\documentclass{article}
\usepackage{fullpage}
\usepackage{listings}
\usepackage{graphicx}
\usepackage[sectionbib,authoryear]{natbib}

\begin{document}
\title{Using Sequence Ensembles for Seeding Alignments of 
MinION Sequencing Data}
\author{Rastislav Rabatin\and Bro\v{n}a Brejov\'a \and Tom\'a\v{s} Vina\v{r}}
\date{Faculty of Mathematics, Physics and Informatics,
           Comenius University,\\ Mlynsk\'a dolina, 842 48 Bratislava, Slovakia}
\maketitle

\begin{abstract}
Oxford Nanopore MinION sequencer is currently the smallest
sequencing device available. While being able to produce very
long reads (reads of up to 100~kbp were reported), it is prone to
high sequencing error rates of up to 30\%. Since most of these
errors are insertions or deletions, it is very difficult to adapt
popular seed-based algorithms designed for aligning 
data sets with much lower error rates.

Base calling of MinION reads is typically done using hidden
Markov models. In this paper, we propose to represent each sequencing
read by an ensemble of sequences sampled from such a probabilistic
model. This approach can improve the sensitivity and false positive rate 
of seeding an alignment compared to using a single representative base
call sequence for each read.
\end{abstract}

\section{Introduction and Background}

In this paper, we explore the use of simple $k$-mer seeding strategies for
mapping MinION reads to the reference sequence. MinION base calls can
have up to 30\% error rate which poses a significant challenge for read 
mapping. Instead of a single read sequence as a query, we propose to
use an ensemble of sequences sampled from a hidden Markov model used
for base calling. With such ensemble of sequences representing
alternative predictions of the true read sequence, we were able to
design a simple seed that
allows for $99.9\%$ sensitivity with a very small number of false
positives on real data.

MinION is currently the smallest and most portable sequencing machine
available. Besides the small size, the advantage of the technology is
its ability to sequence very long reads (reads as long as 100~kbp were
reported). To sequence DNA, MinION uses measurements of electric
current as a single stranded fragment of DNA passes through a
nanopore. The electric current depends mostly on the
context of $k$ bases of DNA passing through the pore at the time of
the measurement. As the DNA fragment moves through the pore, this
context changes and measurements change accordingly.

The raw measurements are first processed by MinKnow software from
Oxford Nanopore. MinKnow uses an on-line algorithm to split raw
measurements into \emph{events}, where each event would ideally
correspond to a single-base shift of the DNA through the pore. Each
event is characterized by the mean and the variance of the
corresponding raw measurements. This sequence of events is then
uploaded to a cloud-based service Metrichor for base calling.

Exact details of the algorithms behind MinKnow and Metrichor are not
disclosed by Oxford Nanopore. However, the whole process is naturally
modelled by a hidden Markov model (HMM) \citep[Chapter 3]{Durbin1998} 
with hidden states
corresponding to $k$-mers of underlying DNA sequence, where
observations would represent mean values of the current in each event.
In fact, data provided by Oxford Nanopore include
parameters of such a model. Open-source base caller based on this
idea was recently implemented by \citet{David2016} in Nanocall
software, with accuracy similar to Metrichor. To decode the sequence
of observations, Nanocall uses the standard Viterbi algorithm 
\citep{Viterbi1967}
for finding the most probable state path. Another open-source base
caller DeepNano uses instead recurrent neural networks
\citep{Boza2016}.

Base calls produced by the Viterbi algorithm contain many errors
\citep{David2016,Boza2016}; a typical error rate would be around 30\%,
dominated mostly by insertions and deletions. With these
characteristics, even basic tasks, such as mapping the reads to the
corresponding reference sequence, become a challenge.

Currently, two general-purpose aligners are used in the community to
map MinION reads: BWA-MEM \citep{Li2010}, and LAST \citep{Kielbasa2011}. 
Both of these
tools follow a general seed-and-extend paradigm, well-known from BLAST
\citep{Altschul1990}.  First, they build an index of one of the sequences (e.g.,
the reference genome), in which they can quickly locate exact matches of
seeds that originated from the query sequence (the read). In this way, 
they identify regions in the reference sequence (called
\emph{hits of a seed}) that are likely to contain an alignment. In
these regions, they perform an extension phase, which will identify
the target alignment.  The \emph{sensitivity} of such tools depends
largely on how likely is a real alignment to contain a hit of a
seed (without the hit, the extension phase is not triggered, and the
alignment is not identified). On the other hand, the \emph{running time}
depends on how many false hits will trigger unnecessary extensions.

The original BLAST \citep{Altschul1990} used 11 consecutive matches as a
seed. Consecutive matches of a fixed length are very easy to index
by standard hashing techniques. For mapping of sequences with
a small number of errors to corresponding reference sequences, longer
seeds were used, offering more specificity (e.g., BLAT
\citep{Kent2002}). BWA-MEM and LAST use variable seed lengths, indexed with 
FM-index \citep{Ferragina2000} or suffix arrays \citep{Manber1993}; by extending seeds to the
point of only a few occurrences, one can avoid most costly false
positives. The adjustments for MinION reads in case of BWA-MEM and
LAST include lowering the minimum length of a seed to be considered as
a valid hit, and changes that make the extension phase less
stringent. GraphMap tool \citep{Sovic2016}, specifically targeting 
MinION data, has been recently
released. GraphMap uses seeds allowing insertions and deletions in the
context of a complex multi-step algorithm that goes well beyond a simple
seed-and-extend framework.

All of these tools consider Metrichor base calls, equivalent to the
most probable state path in the HMM, as the query sequence.  Our goal
in this paper was to explore sub-optimal decodings of the HMM and
attempt to solve the challenges imposed by MinION reads by using an
ensemble of sub-optimal sequences instead of a single DNA sequence. To
this end, we have implemented a sampling algorithm (see,
e.g. \citet{Cawley2003}) that can generate samples from the posterior
distribution of state paths given the sequence of observations. We
adapt common seeding strategies to such ensembles of sequences and show
that on real data we can find a seed that is easy to index, 99.9\%
sensitive, and yields only a small number of false positives that
would trigger extension phase unnecessarily.

Sampling of MinION base calls was also considered by \citet{Szalay2015}.
They use sampling from the base calling HMM to arrive at the correct
consensus sequence for an alignment of multiple reads. Due to the
nature of errors in MinION reads and availability of a reasonable
probabilistic model, we consider sampling strategies to be a promising
alternative in many applications of MinION.

\section{HMM for Sampling MinION Base Calls}
\label{sec:model}

\def\state#1{S_{\mbox{\scriptsize #1}}}

Both Oxford Nanopore Metrichor base caller and recently released
open-source base caller Nanocall \citep{David2016} use hidden
Markov models. Briefly, each hidden state of the HMM represents one $k$-mer
passing through the pore, and the emission of the state is the value 
of the electric current. Actual measurements of current 
provided by the device with high sampling rate are segmented 
by the MinKnow software into discrete \emph{events}, each corresponding to 
the shift of the DNA sequence through the pore by a single base. 
The base callers then use an HMM to obtain the sequence of 
hidden states given the sequence of events from the MinION read.

\paragraph{Definition of the model.}
Our HMM follows the same general idea. The set of states of our
HMM is composed of all $k$-mers (we denote state for a $k$-mer $x$ by
$S_x$) and the starting state $S_0$. Different versions of MinION
use different values of $k$; in our experiments we have used a data set
with $k=5$, while the newer chemistry uses $k=6$.

Emission of state $S_x$ is represented as a continuous random
variable. The probability of observing a measurement $e$
for a $k$-mer $x$ is given as
\begin{equation}
  \Pr(e\,|\,x) \sim \mathcal{N} (\mathit{scale}\cdot \mu_x + \mathit{shift},
                                  \sigma_x\cdot \mathit{var}),
\end{equation}
where $\mathcal{N}(\mu,\sigma)$ is a normal distribution with mean $\mu$ and
standard deviation $\sigma$. Parameters
$\mu_x$, $\sigma_x$ (specific for each version of the chemistry and
each $k$-mer $x$), and $\mathit{scale}$, $\mathit{shift}$,
$\mathit{var}$ (scaling parameters specific for each read) are
provided by Oxford Nanopore and can be obtained from the FAST5 file
containing each read. Starting state $S_0$ is silent.

Under ideal conditions, each event corresponds to a shift by a single base
in the DNA sequence. This corresponds to four outgoing transitions
from each state $S_x$ to state $S_y$, where $x$ and $y$ overlap by
exactly $k-1$ bases (i.e., $\state{AACTG}$ has transitions to
states $\state{ACTGA}$, $\state{ACTGC}$, $\state{ACTGG}$, and
$\state{ACTGT}$). This organization closely resembles de Bruijn graphs
commonly used in sequence assembly \citep{Pevzner2001}. 
All four transitions have have an
equal probability. From the starting state $S_0$, we have a transition
to each possible $S_x$ with equal probability $1/4^k$.

Segmentation of raw measurements into events is  
known to be error prone. In particular,
two events with similar measurements can be fused together, or a
single event can be split artificially into multiple events. 
Thus the assumption that each event corresponds to a single-nucleotide
shift is unrealistic. To account for this
fact, we have introduced additional transitions in our model.

First, we have added a self-transition (so called \emph{split
  transition}) to each state, which models splitting of a single true event
into multiple  predicted events. Second, we have also added so called \emph{skip
  transitions} between all pairs of states $S_x$ and $S_y$, which
would correspond to shifts of the $k$-mer by up to $k$ bases instead
of one.

The transitions probabilities for split and skip transitions are not
provided by MinION. We have estimated these parameters
directly from the data, as outlined in Section \ref{sec:training}.
Alternatingly, we could employ a more elaborate error model for 
Oxford Nanopore event segmentation process.

\paragraph{Inferrence in the model.}

A traditional way of decoding HMMs is by finding the most probable
sequence of states by the Viterbi algorithm \citep{Viterbi1967}, which is the
approach taken both by Metrichor and Nanocall. The resulting sequence
of states (which is, in fact, a sequence of $k$-mers corresponding to
individual events) can be translated into the DNA sequence. In most
cases, the neighbouring $k$-mers in the sequence should be shifted by
one, and thus each state in the sequence should introduce one
additional base of the DNA sequence. However, split and skip
transitions may introduce between 0 and $k$ bases for each event. In
these cases, the result is not necessarily unique: for example state
sequence $\state{ACTCTC}\state{CTCTCA}$ could correspond to one of the
sequences ACTCTCA, ACTCTCTCA, ACTCTCTCTCA, or even ACTCTCCTCTCA. 
We have decided to adopt the shortest possible
interpretation, as is done in Nanocall.

Since the base calls produced by the Viterbi algorithm contain
many errors, we have decided to explore the use of an ensemble of
alternative sequences instead of a single base call sequence. To this end,
we have implemented the stochastic traceback algorithm (see,
e.g. \citet{Cawley2003}) to generate samples from the posterior
distribution of state paths given the sequence of observations. 

Briefly, for a given sequence of observations $e_1e_2\dots e_n$, the
algorithm starts by computing \emph{forward probabilities}, where
probability $F(i,s)$ is the probability of generating first $i$
observations $e_1e_2\dots e_i$, and ending in state $s$ \citep{Rabiner1989}. The last
state $s_n$ of the path is sampled proportionally to the probabilities
$F(n,s_n)$. When state $s_n$ is fixed, we can sample state $s_{n-1}$
proportionally to $F(n-1,s_{n-1})\cdot t_{s_{n-1},s_n}$, where
$t_{s,s'}$ is the transition probability from state $s$ to state
$s'$. This can be continued, until we sample the complete path
$s_1\dots s_n$. The running time of the algorithm is $O(nm^2)$,
where $n$ is the length of the sequence and $m$ is the number of states. 
Forward
probabilities need to be computed only once if multiple samples are
required for the same read.

\begin{figure}[t]
\includegraphics[width=\textwidth]{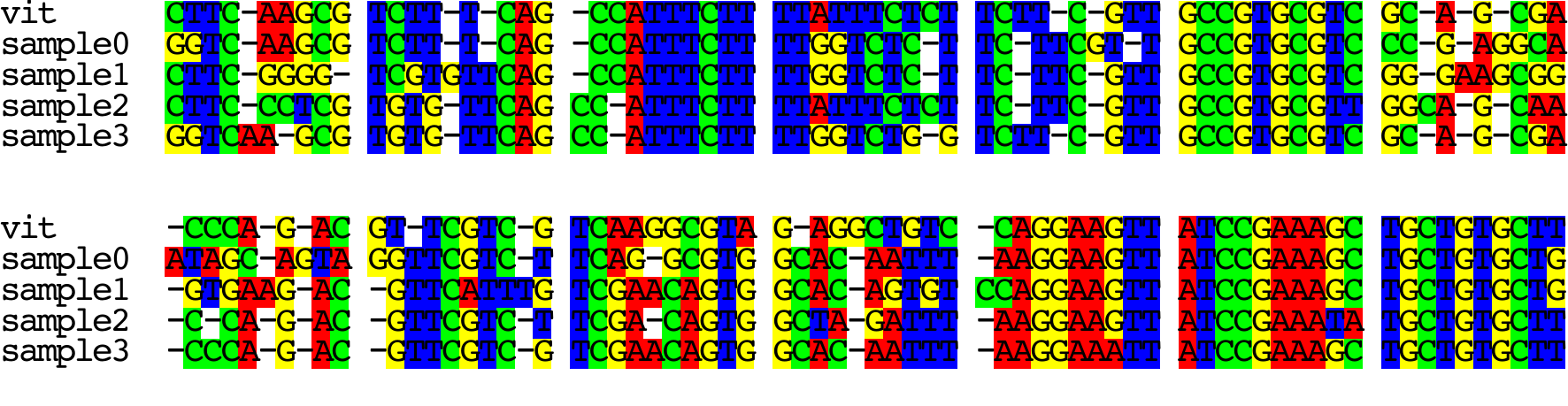}
\caption{{\bf Example of base calling samples from a MinION read.} The first
  line corresponds to the Viterbi base call, other lines correspond to
  four samples from the posterior distribution defined by the HMM. Base calls
  were aligned according to events in the sequence of observations.}
\label{fig:samples}
\end{figure}

Figure \ref{fig:samples} illustrates typical differences between
individual samples. Note that the samples are almost
identical in some regions, 
but these conserved regions are interspersed by regions with very high uncertainty.
This is a typical pattern for MinION data.

\section{Experiments}
\label{sec:training}

\paragraph{Data preprocessing and model training.}
For our experiments, we have used a data set from \emph{E. coli} (strain
MG1655) with accession number ERR968968 produced by the Cold Spring Harbor
Laboratory by using MinION sequencer with SQK-MAP005 kit. For simplicity,
we have only considered template reads (complement reads
from the reverse strand use different model parameters).

To filter out low quality reads, we have mapped Metrichor base calls 
to the reference sequence by BWA-MEM \citep{Li2010} with
\verb'-x ont2d' parameters optimized for mapping Oxford Nanopore
reads.  The reads that did not map to the reference at all were
discarded. We also discarded reads where Metrichor predicted skips
in the event sequence 
longer than two. From the original 27,073 reads, we were left with
25,162 reads.
 
From these reads, we have randomly selected a training set (693 reads)
and a testing set (307 reads). The training set was used to estimate
the transition probabilities in our HMM.  In
particular, we set the probability of each transition to be
proportional to the number of times the transition was observed in the
training data set. We added pseudocount of 1 to avoid zero
transition probabilities for rare transitions.

\begin{figure}[t]
\centerline{\includegraphics[width=0.8\textwidth]{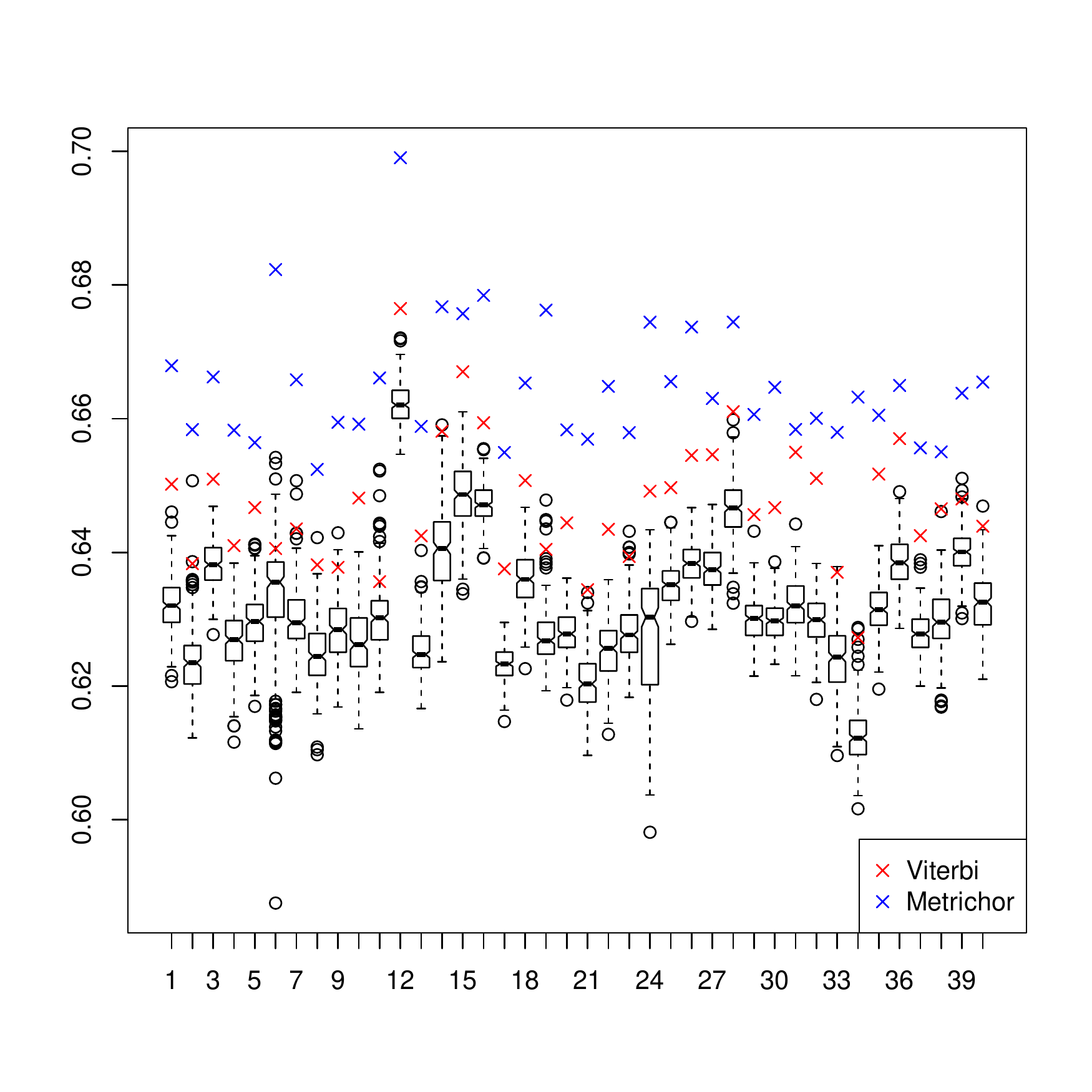}}
\vglue -2\baselineskip
\caption{{\bf Comparison of sequence identity to the reference genome 
for Metrichor base calls, our Viterbi base
    calls, and posterior samples for 40 randomly selected reads.}
  All sequences were aligned to the reference by BWA-MEM.
  Sequence identity of Metrichor, Viterbi, and a box plot distribution
  of sequence identities of 250 samples are shown on the $y$-axis.}
\label{fig:boxplot}
\end{figure}

\paragraph{Preparing testing data.} 
For each sequence in the testing set, we have produced a Viterbi base
call and 250 samples from the posterior distribution as outlined in
Section \ref{sec:model}. Figure \ref{fig:boxplot} shows comparison
of sequence identities of individual base calls to the reference genome.
Note that our
Viterbi base calls are not too different from Metrichor base calls;
slight decrease in the quality of calls is expected due to simplicity
of the model we have used (the decrease in the sequence identity is
similar to that observed by \citet{David2016}). Sampled
sequences have in general lower sequence identity than the Viterbi
base call, as can be expected, since they are mostly sub-optimal paths
through the model. However, the difference from the Viterbi base
call identities is not very high.

\paragraph{Experimental setup and evaluation.}
Our goal is to consider various seeding strategies for seed-and-extend
algorithms similar to BLAST \citep{Altschul1990}. Briefly, a typical
seed-and-extend algorithm first uses an index structure to locate hits
between the query sequence and the target. For example, the most
basic BLAST strategy looks for exact matches of length 11. 
Second, we try to extend each cluster
of hits to a full alignment. The extension phase usually involves dynamic
programming and is therefore time consuming. 

The seed-and-extend algorithms cannot locate alignments that 
do not contain a hit of the seed between the query and
the target sequence.  We call these alignments \emph{false
  negatives}. Note that even a single hit is often sufficient to locate the
whole alignment, depending on the particular extension strategy. 
On the other hand, there will be spurious hits
between random locations which will trigger extension. These spurious
hits often dominate the running time of the alignment algorithm, 
and consequently, we
need to minimize their number. We call such spurious hits \emph{false
positives.}

To evaluate application of various seeding strategies in MinION data,
we will split our data set into \emph{windows}, each corresponding to
500 events. More precisely, for each read all base calls and samples
were aligned to each other based on event boundaries: we have padded
all sequences generated for one event in individual samples by gap
symbols so that they have the same length (see also Figure
\ref{fig:samples}).  The resulting multiple alignment was then split into
windows which will represent our query sequences.

The Viterbi base call sequence from each window was aligned to the
reference genome by LAST software \citep{Kielbasa2011} with parameters
\verb'-q 1 -a 1 -b 1 -T 1'. We have kept only those windows that
aligned to a unique place in the genome, and the alignment covered the
entire length of the window. After this step, we were left with 3192
windows out of 4948. A randomly chosen subset of 143 windows was used
as a validation set for exploring various seeding strategies, and the
remaining 3049 windows were used for final testing. The region of the
reference covered by the alignment to the window is considered to be
the only true alignment of the window to the reference sequence.

When testing an alignment seeding strategy, we try to locate a particular seed
in both the query window and the reference sequence. We represent the 
hit of the seed 
by the coordinates of the left endpoints of the hit
in the query window and in the reference. The seed hit is considered to
be \emph{valid}, if the endpoint in the reference is within the region
covered by the alignment of this particular window and on the correct
strand. The entire window of a read is considered to be a \emph{true
  positive (TP)}, if it contains at least one valid hit; we assume
that the extension algorithm would be able to recover the alignment
within this window starting from this seed. The \emph{sensitivity
  (Sn)} of the seeding strategy is the number of true positives divided
by the total number of windows. 

Many seed hits are invalid, and they contribute to the false positive
rate.  Often we see clusters of seeds with very similar coordinates in
both reference and the query window. Presumably the extension
algorithm would be called only once for each such cluster. Therefore,
we compute the number of \emph{false positives} by greedily selecting
one seed from each cluster so that each seed differs in both
coordinates by at most 10 from a selected read.

\paragraph{Simple seeding strategies.} The most simple seeding
strategy is to consider $k$ consecutive exact matches as a hit.  The
traditional approach would create an index of all $k$-mers in the
reference genome and then scan all $k$-mers in the query windows. Each
cluster of matching $k$-mers would trigger the extension phase.  

\begin{figure}[t]
\begin{minipage}[b]{0.6\textwidth}
~\\[-\baselineskip]
\includegraphics[width=0.98\textwidth]{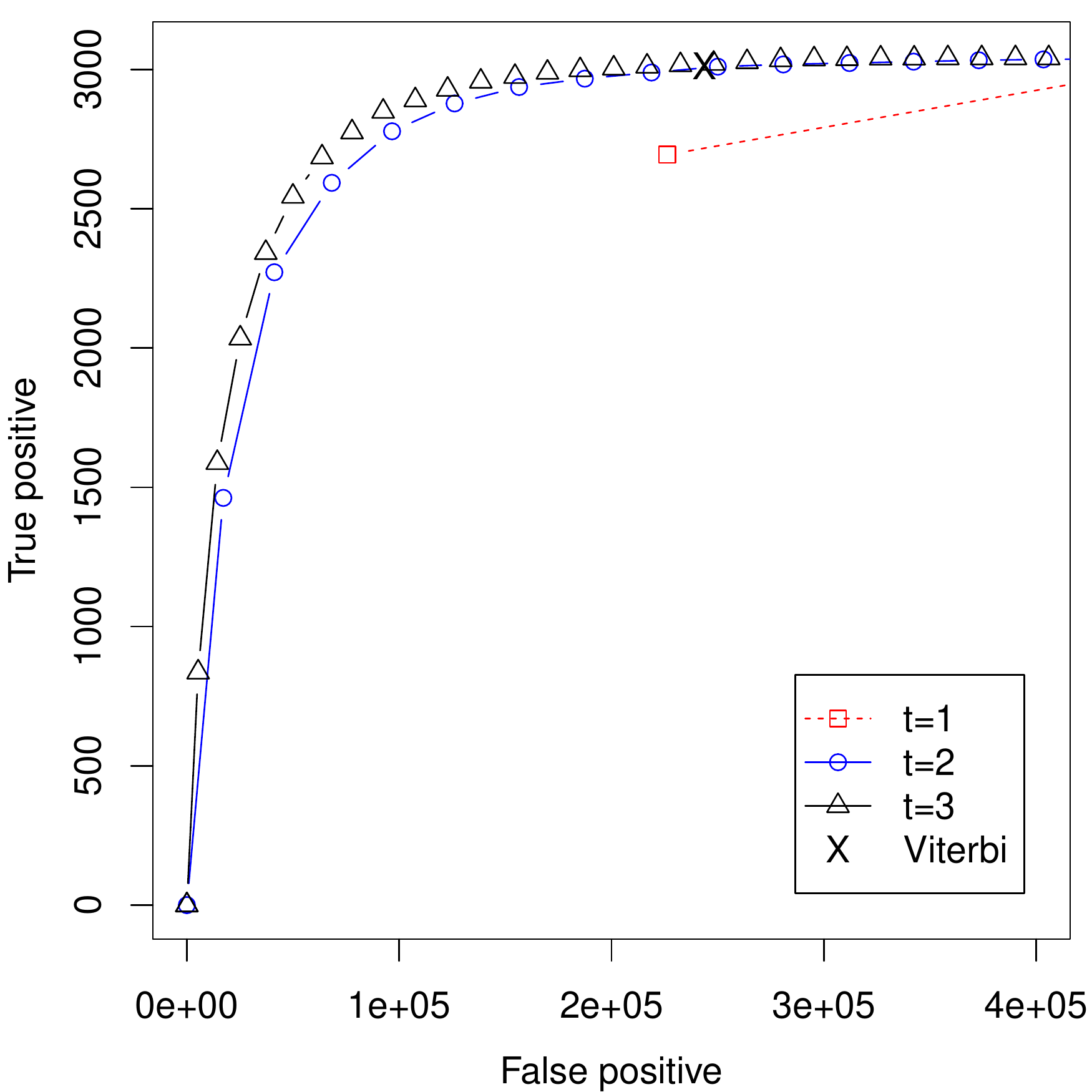}
\end{minipage}
\hfill
\begin{tabular}[b]{l@{\quad}r@{\quad}r}
\hline
Method & Sn & FP \\
\hline
Viterbi      &   0.988 &  243796 \\[1mm]
$t=1, n=  2$ &   0.974 &  433890 \\
$t=1, n=  3$ &   0.990 &  625782 \\
$t=1, n=  8$ &   0.999 & 1456049 \\[1mm]
$t=2, n=  7$ &   0.963 &  156472 \\
$t=2, n= 11$ &   0.990 &  280912 \\
$t=2, n= 26$ &   0.999 &  727678 \\[1mm]
$t=3, n= 12$ &   0.960 &  122830 \\
$t=3, n= 18$ &   0.988 &  216743 \\
$t=3, n= 38$ &   0.999 &  529692 \\[1mm]
$t=4, n= 17$ &   0.957 &  110197 \\
$t=4, n= 25$ &   0.989 &  192648 \\
$t=4, n= 53$ &   0.999 &  488198 \\[1mm]
$t=5, n= 22$ &   0.956 &  103666 \\
$t=5, n= 34$ &   0.988 &  196604 \\
$t=5, n= 72$ &   0.999 &  497944 \\
\hline
\end{tabular}
\caption{{\bf
Performance of our approach compared to the Viterbi base calls for 
seeding with a single $13$-mer.} 
The $x$-axis of the plot is the total number of false positives on the whole
testing set; the $y$-axis is the number of true positives out of 
3049 windows in total. 
Performance of the Viterbi base calls is shown by the black $X$. 
Three lines show our approach for different values of
threshold $t$ for filtering $k$-mers. Each point on the line represents
performance for a particular number of samples $n=1,2,\dots$.
The table shows sensitivity and the number of false positives for
selected values of $n$ and $t$. In particular for each $t$, we show
the smallest value of $n$ achieving sensitivity at least $95\%$,
the sensitivity of the Viterbi algorithm, and sensitivity at least
$99.9\%$.
}
\label{fig:res13}
\end{figure}

For example, if we consider the Viterbi base calls and use 13 exact
consecutive matches as a seed, we will be able to map $98.8\%$ windows
to the correct region of the reference (see
Fig. \ref{fig:res13}), but we will also incur a
substantial number of false positives (more than 240,000 or about 80
per each query window).

Our strategy of using sampling instead of Viterbi base calls works as
follows.  In the basic version of our approach (see $t=1$ in
Fig. \ref{fig:res13}), we consider $k$-mers from $n$ different samples from
the HMM for each position in the read window. Each $k$-mer can potentially 
form the seed triggering the extension phase of the
alignment. To match the sensitivity of the Viterbi base calls for $k=13$, 
we need
only $n=3$ samples. One advantage of our approach is that we can
increase sensitivity by increasing the number of samples $n$. For
example with 8 samples we reach 99.9\% sensitivity and with 14 samples
100\% sensitivity. The cost for this very high sensitivity is a high
false positive rate; even with 3 samples we have about 2.6$\times$
more false positives than the Viterbi.

To improve the false positive rate, we use a simple prefiltering step:
at each position in the window we consider only those $k$-mers that
appear in at least $t$ different samples.
Assuming that the true sequence has a high posterior
probability in the model, we expect that it will occur in many
samples, whereas most other variants would occur rarely and thus
be filtered out. Indeed, for $t=4$ we can reach the sensitivity of the
Viterbi algorithm with about 20\% reduction in the false positive
rate, using $n=25$ samples.

\paragraph{Multiple seed hits to trigger extension.}
We have also considered a more complex seeding strategy, where we
first find matching $10$-mers and then we join them into \emph{chains}
of length 3. This technique has previously proved to be very 
effective for regular
alignment tasks \citep{Altschul1997}. 

Matching $10$-mers 
in the chain are required to have increasing coordinates in both
the read and the reference sequence, and the distance between starts of
adjacent seeds in the chain must be at least 10 and at most 50
in both sequences. However, the distances of the two $10$-mer matches 
in the two sequences may differ, accommodating indels 
in the intervening regions. The
entire chain is then again represented by its leftmost point in both the
read and the reference for the purpose of determining if it is valid.

\begin{figure}[t]
\begin{minipage}[b]{0.6\textwidth}
~\\[-\baselineskip]
\includegraphics[width=0.98\textwidth]{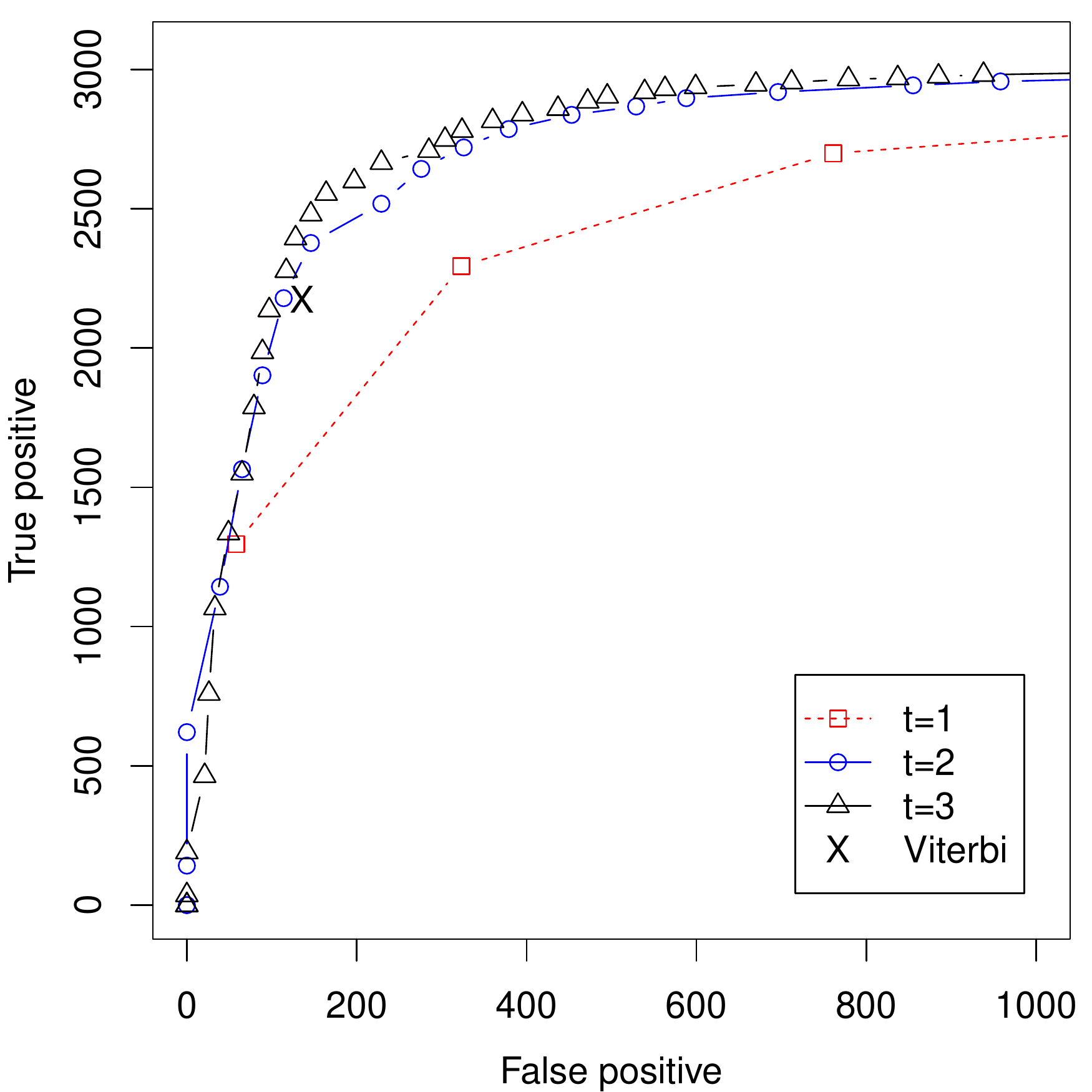}
\end{minipage}
\hfill
\begin{tabular}[b]{l@{\quad}r@{\quad}r}
\hline
Method & Sn & FP \\
\hline
Viterbi        &   0.713 &     136 \\[1mm]
$t=1, n=  2$ &   0.752 &     323 \\
$t=1, n=  5$ &   0.967 &    2412 \\
$t=1, n= 13$ &   0.999 &   18249 \\[1mm]
$t=2, n=  7$ &   0.715 &     114 \\
$t=2, n= 15$ &   0.950 &     588 \\
$t=2, n= 43$ &   0.999 &    6407 \\[1mm]
$t=3, n= 13$ &   0.747 &     117 \\
$t=3, n= 26$ &   0.952 &     495 \\
$t=3, n= 80$ &   0.999 &    6180 \\[1mm]
$t=4, n= 18$ &   0.714 &      89 \\
$t=4, n= 39$ &   0.950 &     517 \\
$t=4, n=139$ &   0.999 &    8877 \\[1mm]
$t=5, n= 24$ &   0.721 &     139 \\
$t=5, n= 51$ &   0.954 &     483 \\
$t=5, n=165$ &   0.999 &    7003 \\
\hline
\end{tabular}
\caption{
{\bf Performance of our approach compared to the Viterbi base calls for
seeding with a chain of three $10$-mers,} each in distance at most 50
from the previous one. The plot and the table have the same form as in Figure 
\ref{fig:res13}.
}
\label{fig:res10}
\end{figure}

As we see in Figure \ref{fig:res10}, this seeding
strategy is too stringent for the Viterbi base calls, achieving only
71.3\% sensitivity. On the other hand, false positives are extremely
rare, totaling only 136 in all 3049 windows.

When using samples, different $10$-mer matches in the chain may come from
different samples. The chaining of weaker seeds helps to accommodate
regions with high uncertainty and many indels present in the MinION
data.  Some settings of our
strategy can achieve the same sensitivity as the Viterbi algorithm 
with even lower false
positives, but more importantly, by considering more samples, we can
increase sensitivity while keeping the false positives quite low.
For example, for $t=2$ and $n=43$ our strategy can reach 99.9\%
sensitivity with only 6407 false positives.

\section{Conclusions and Future Work}

In this paper, we have examined the problem of mapping MinION
sequencing reads to the reference sequence. The error rate of MinION
reads is very high (approx. 30\%), with many insertions and
deletions. Consequently the standard sequence alignment techniques 
do not achieve sufficient sensitivity for mapping reads.

Instead of representing the read by 
a single base called sequence, we have proposed to
use an ensemble of sequences generated from the posterior distribution
defined by the HMM capturing the properties of MinION sequencing
process. We have adapted the standard $k$-mer based techniques for alignment
seeding to ensembles of sequences and identified a seed (three 10-mer
hits spaced by at most 50 bases) that in our experiments achieved
99.9\% sensitivity with an extremely small number of false positives.
With such a low false positive rate, we could investigate more precise
(and slower) algorithms for the extension phase, which will be the next
important step towards sensitive alignment of MinION reads. 

An obvious extension of our approach 
would be to consider spaced seeds \citep{Ma2002}.  Our
experiments suggest that a typical MinION read consists of short
regions of high-confidence sequence (often under 15bp) interspersed
with regions of high uncertainty with many indels.  The spaced seeds
would have to target mainly these high-confidence regions, however,
these regions seem to be too short to admit complex seeds of any
significant weight. One possibility would be to build a probabilistic
model capturing high-confidence and high-uncertainty regions and
transitions between them, and attempt to design an optimized spaced
seeds, for example by techniques suggested by \citet{Brejova2004}.  

Another option would be to use seeds that also allow indels at do not
care positions. These types of seeds were successfully used by
\citet{Sovic2016} for MinION read mapping, but the overall algorithm
was much more complicated than a simple seed-and-extend. Moreover,
these types of seeds are much more difficult to index than continuous
or spaced seeds and we believe, that our sampling approach together
with multiple chained seed hits provides an elegant answer to the problem.

In this work, we have only considered 1D reads from MinION. However,
MinION attempts to read both strands of the DNA and then combine the
readouts in postprocessing to a single sequence (these are called 2D
reads). Since 2D reads are much more accurate (typical error rate is
about 15\%), most of the researchers using MinION suggest throwing out
1D reads and only use 2D reads in further analysis. 

This has two problems. First, usually there is about $4-5\times$ more
1D reads than 2D reads that pass Metrichor base calling
procedure \citep{Ip2015}. Thus we are throwing out most of the data.
Second, recently people have started to use MinION in applications such
as monitoring disease outbreaks \citep{Quick2016}. In these
applications, reads are analyzed on-the-fly as they are produced, and 
 we cannot rely on postprocessing of base calls.

In future work, we would like to investigate the seed-and-extend
approaches to read-to-read alignment. With our sampling approach, we
would not need to commit to a single interpretation of either of the
sequences, potentially increasing the sensitivity of detecting overlaps
between reads in a given data set. Sensitive read-to-read alignment is
essential for \emph{de novo} assembly.

\paragraph{Acknowledgements.} This research was funded by APVV
grant APVV-14-0253 and VEGA grants 1/0719/14 (TV) and 1/0684/16 (BB).

\bibliographystyle{apalike} 
\bibliography{main}

\begin{thebibliography}{}

\bibitem[Altschul et~al., 1990]{Altschul1990}
Altschul, S.~F., Gish, W., Miller, W., Myers, E.~W., and Lipman, D.~J. (1990).
\newblock {Basic local alignment search tool}.
\newblock {\em Journal of Molecular Biology}, 215(3):403--410.

\bibitem[Altschul et~al., 1997]{Altschul1997}
Altschul, S.~F., Madden, T.~L., Schaffer, A.~A., Zhang, J., Zhang, Z., Miller,
  W., and Lipman, D.~J. (1997).
\newblock {Gapped BLAST and PSI-BLAST: a new generation of protein database
  search programs}.
\newblock {\em Nucleic Acids Research}, 25(17):3389--3392.

\bibitem[Bo{\v{z}}a et~al., 2016]{Boza2016}
Bo{\v{z}}a, V., Brejov{\'a}, B., and Vina{\v{r}}, T. (2016).
\newblock {DeepNano}: Deep recurrent neural networks for base calling in
  {MinION} nanopore reads.
\newblock Technical Report arXiv:1603.09195, arXiv.org.

\bibitem[Brejova et~al., 2004]{Brejova2004}
Brejova, B., Brown, D.~G., and Vinar, T. (2004).
\newblock {Optimal spaced seeds for homologous coding regions}.
\newblock {\em Journal of Bioinformatics and Computational Biology},
  1(4):595--610.

\bibitem[Cawley and Pachter, 2003]{Cawley2003}
Cawley, S.~L. and Pachter, L. (2003).
\newblock {HMM sampling and applications to gene finding and alternative
  splicing}.
\newblock {\em Bioinformatics}, 19(S2):ii36--41.

\bibitem[David et~al., 2016]{David2016}
David, M., Dursi, L.~J., Yao, D., Boutros, P.~C., and Simpson, J.~T. (2016).
\newblock Nanocall: An open source basecaller for {Oxford Nanopore} sequencing
  data.
\newblock Technical Report bioRxiv:046086, Cold Spring Harbor Laboratory.

\bibitem[Durbin et~al., 1998]{Durbin1998}
Durbin, R., Eddy, S.~R., Krogh, A., and Mitchison, G. (1998).
\newblock {\em Biological Sequence Analysis: Probabilistic Models of Proteins
  and Nucleic Acids}.
\newblock Cambridge University Press.

\bibitem[Ferragina and Manzini, 2000]{Ferragina2000}
Ferragina, P. and Manzini, G. (2000).
\newblock Opportunistic data structures with applications.
\newblock In {\em Foundations of Computer Science (FOCS)}, pages 390--398.
  IEEE.

\bibitem[Ip et~al., 2015]{Ip2015}
Ip, C. L.~C., Loose, M., Tyson, J.~R., {de Cesare}, M., Brown, B.~L., Jain, M.,
  Leggett, R.~M., Eccles, D.~A., Zalunin, V., Urban, J.~M., Piazza, P., Bowden,
  R.~J., Paten, B., Mwaigwisya, S., Batty, E.~M., Simpson, J.~T., Snutch,
  T.~P., Birney, E., Buck, D., Goodwin, S., Jansen, H.~J., O'Grady, J., and
  Olsen, H.~E. (2015).
\newblock {MinION Analysis and Reference Consortium: Phase 1 data release and
  analysis}.
\newblock {\em F1000Research}, 4:1075.

\bibitem[Kent, 2002]{Kent2002}
Kent, W.~J. (2002).
\newblock {BLAT--the BLAST-like alignment tool}.
\newblock {\em Genome Research}, 12(4):656--664.

\bibitem[Kielbasa et~al., 2011]{Kielbasa2011}
Kielbasa, S.~M., Wan, R., Sato, K., Horton, P., and Frith, M.~C. (2011).
\newblock {Adaptive seeds tame genomic sequence comparison}.
\newblock {\em Genome Research}, 21(3):487--493.

\bibitem[Li and Durbin, 2010]{Li2010}
Li, H. and Durbin, R. (2010).
\newblock {Fast and accurate long-read alignment with Burrows-Wheeler
  transform}.
\newblock {\em Bioinformatics}, 26(5):589--595.

\bibitem[Ma et~al., 2002]{Ma2002}
Ma, B., Tromp, J., and Li, M. (2002).
\newblock {PatternHunter: faster and more sensitive homology search}.
\newblock {\em Bioinformatics}, 18(3):440--445.

\bibitem[Manber and Myers, 1993]{Manber1993}
Manber, U. and Myers, G. (1993).
\newblock Suffix arrays: a new method for on-line string searches.
\newblock {\em SIAM Journal on Computing}, 22(5):935--948.

\bibitem[Pevzner et~al., 2001]{Pevzner2001}
Pevzner, P.~A., Tang, H., and Waterman, M.~S. (2001).
\newblock {An Eulerian path approach to DNA fragment assembly}.
\newblock {\em Proceedings of the National Academy of Sciences of the USA},
  98(17):9748--9753.

\bibitem[Quick et~al., 2016]{Quick2016}
Quick, J. et~al. (2016).
\newblock {Real-time, portable genome sequencing for Ebola surveillance}.
\newblock {\em Nature}, 530(7589):228--232.

\bibitem[Rabiner, 1989]{Rabiner1989}
Rabiner, L.~R. (1989).
\newblock A tutorial on hidden {M}arkov models and selected applications in
  speech recognition.
\newblock {\em Proceedings of the IEEE}, 77(2):257--286.

\bibitem[Sovic et~al., 2016]{Sovic2016}
Sovic, I., Sikic, M., Wilm, A., Fenlon, S.~N., Chen, S., and Nagarajan, N.
  (2016).
\newblock {Fast and sensitive mapping of nanopore sequencing reads with
  GraphMap}.
\newblock {\em Nature Communications}, 7:11307.

\bibitem[Szalay and Golovchenko, 2015]{Szalay2015}
Szalay, T. and Golovchenko, J.~A. (2015).
\newblock {De novo sequencing and variant calling with nanopores using
  PoreSeq}.
\newblock {\em Nature Biotechnology}, 33(10):1087--1091.

\bibitem[Viterbi, 1967]{Viterbi1967}
Viterbi, A.~J. (1967).
\newblock Error bounds for convolutional codes and an asymptotically optimum
  decoding algorithm.
\newblock {\em IEEE Transactions on Information Theory}, 13(2):260--269.

\end{thebibliography}

\end{document}